\documentclass[useAMS,onecolumn]{mn2e}
\usepackage{epsfig}
\title[SPM X-ray binaries]{High proper motion X-ray binaries from the Yale Southern Proper Motion Survey}

\author[Maccarone et al.]{Thomas
  J. Maccarone$^{1}$\thanks{email:thomas.maccarone@ttu.edu},  Terrence
  M. Girard$^2$ \& Dana I. Casetti-Dinescu$^{3,2}$\\
$^{1}$ Department of Physics, Texas Tech University, Lubbock TX, 79409, USA \\
$^2$ Department of Astronomy, Yale University, 260 Whitney Avenue, New Haven, CT, 06511 USA\\
$^3$ Department of Physics, Southern Connecticut State University, 501
Crescent Street, New Haven CT, 06515, USA\\}
\begin{document}
\def\ltsim{\mathrel{\rlap{\lower 3pt\hbox{$\sim$}}
        \raise 2.0pt\hbox{$<$}}}
\def\gtsim{\mathrel{\rlap{\lower 3pt\hbox{$\sim$}}
        \raise 2.0pt\hbox{$>$}}}

\date{}

\pagerange{\pageref{firstpage}--\pageref{lastpage}} \pubyear{}

\maketitle

\label{firstpage}

\begin{abstract}
We discuss the results of cross-correlating catalogs of bright X-ray
binaries with the Yale Southern Proper Motion catalog (version 4.0).
Several objects already known to have large proper motions from
Hipparcos are recovered.  Two additional objects are found which show
substantial proper motions, both of which are unusual in their X-ray
properties.  One is IGR J17544-2619, one of the supergiant fast X-ray
transients.  Assuming the quoted distances in the literature for this
source of about 3 kpc are correct, this system has a peculiar velocity
of about 275 km/sec -- greater than the velocity of a Keplerian orbit
at its location of the Galaxy, and in line with the expectations
formed from suggestions that the supergiant fast X-ray transients
should be highly eccentric.  We discuss the possibility that these
objects may help explain the existence of short gamma-ray bursts
outside the central regions of galaxies. The other is the source
2A~1822-371, which is a member of the small class of objects which are
low mass X-ray binaries and long (i.e. $>$ 100 millisecond) X-ray
pulsars.  This system also shows both an anomalously high X-ray
luminosity and a large orbital period derivative for a system with its
orbital period, and some possible indications of an eccentric orbit.
A coherent picture can be developed by adding in the proper motion
information in which this system formed in the Perseus spiral arm of
the Galaxy about 3 Myr ago, and retains a slightly eccentric orbit
which leads to enhanced mass transfer.

\end{abstract}


\section{Introduction}

The mechanisms by which supernovae explode have been a topic of recent
debate (e.g. Belczynski et al. 2012).  X-ray binaries represent one of
the few classes of systems in the Universe in which the end results of
supernova explosions can be explored, and the mass distribution of
compact objects has been proposed as a key test of the supernova
mechanism.  The systems can be readily identified from their X-ray
emission, allowing them to be selected easily for further study.  The
masses of the compact objects can be estimated from the radial
velocity motions of the donor stars, along with estimates of their
inclination angles from eclipses, in an ideal case, or ellipsoidal
modulations (see Cantrell et al. 2010 for the state of the art) or
polarimetry (Brown et al. 1978; e.g. Dolan \& Tapia 1989).
Astrometric wobble has been seen in Cygnus X-1 (Reid et al. 2011), but
not yet at a level where it could be used to make precise measurements
of a system inclination angle.  Spaced-based astrometric surveys such
as Gaia or the proposed SIM mission do have the potential to measure
the inclination angles of wide, nearby X-ray binaries (e.g. Tomsick et
al. 2009; Tomsick \& Muterspaugh 2010).

Along with the mass distribution of compact objects, it is interesting
to understand their space velocity distribution.  A few different
mechanisms have been proposed for giving X-ray binaries (and, indeed,
isolated neutron stars) space velocities far in excess of those of
their massive star progenitors.  The loss of mass in supernovae on
timescales fast compared with the orbital period means that the mass
lost in the supernova will carry away linear momentum relative to the
center of mass of the binary system, meaning that the binary itself
will receive an equal an opposite momentum kick (see e.g. Blaauw
1961).  These kicks are sometimes referred to as Blaauw kicks, or as
symmetric kicks, since they are symmetric about the supernova
progenitor's center of mass.

A different class of supernova kicks have also been proposed --
asymmetric kicks, which are, as suggested from the name, asymmetric
about the center of mass of the supernova progenitor.  A variety of
theoretical mechanisms exist for producing such kicks -- hydrodynamic
kicks, due to actual asymmetries in the mass ejection (e.g. Burrows \&
Hayes 1996), interactions between neutrinos and a strong magnetic
field (e.g. Arras \& Lai 1999), and electromagnetic (e.g. Harrison \&
Tademaru 1975) -- see Lai et al. (2000) for a review.

From relatively early on in the study of pulsars, it was found that
they had typical space velocities of order 100 km/sec or more (Trimble
\& Rees 1971; Lyne et al. 1982) -- much larger than the space
velocities of the OB stars which are their progenitors.  In some
cases, these high space velocities might be explained as a ``Blauuw
kick''.  A variety of lines of evidence, mostly from observations of
X-ray binaries, indicate that some additional kick component is likely
to be present (see Podsiadlowski et al. 2005 for a discussion of the
array of evidence).  In principle, hypervelocity stars can also be
produced in dynamical interactions -- for example with the
supermassive black hole in the Galactic Center (e.g. Hills 1988; Brown
et al. 2006; Rossi et al. 2013), although the velocities will then
always be radial, away from the Galactic Center.

Understanding the space velocities of X-ray binaries gives one route
to estimating the kick distribution of neutron stars at birth.
Understanding the kick velocities in X-ray binaries is important for
understanding their formation and evolution, the retention of neutron
stars in globular clusters (e.g. Pfahl et al. 2002; Smits et
al. 2006), and for making connections between the populations of high
mass X-ray binaries and gravitational wave sources (e.g. Belczynski et
al. 2011, 2013).

In this paper, we report on the discovery of large proper motions from
two X-ray binaries, the low mass X-ray binary 2A~1822-371, and the
supergiant fast X-ray transient IGR~J17544-2619.  In both cases, the
high energy properties of the systems are unusual, and large space
velocities may help to understand the systems.

\section{The data}

We work here with several public catalogs, the Yale Southern Proper
Motion (SPM) Catalog, version 4.0 (Girard et al. 2011), and the catalogs of
low mass X-ray binaries (these can come from burst oscillations --
Strohmayer et al. 1996; or from accreting millisecond pulsars --
Wijnands et al. 1998; see Liu et al. 2007 for a compilation) and high
mass X-ray binaries (Liu et al. 2006) which represent the updates to
the van Paradijs catalog (van Paradijs et al. 1995).  We use the CDS
X-match facility to cross-correlate the catalogs, taking all matches
within 5'' in order to allow for the fact that some of the X-ray
positions for some sources will be of poor quality, while also not
taking such a large search radius that the matches will be heavily
dominated by chance superpositions.  In many cases, the positions in
the Liu catalog are not accurate to the sub-arcsecond level, because
the source positions were not known to high precision at the time of
publications of the catalogs -- for this reason we begin by including
some sources whose positions are separated by more than the positional
accuracies of the sources in the SPM catalog, and then filter out the
sources which in hindsight appear to be chance superpositions by
making use of additional information from SIMBAD.  As an extra check
on the accuracy of the proper motions we report here, we have also
considered the same sources' proper motions in the fourth UCAC catalog
(Zacharias et al. 2013), and the cross-calibration between proper
motions in Hipparcos (Perryman et al. 1997) and SPM within 1 degree of
our sources of interest.

\subsection{High mass X-ray binaries}

There are 27 matches within 5'' between the SPM catalog and the HMXB
catalog.  Of these, 22 are within 1'' and 24 are within 1.5'', with
the uncertainties in the catalogued X-ray positions likely to dominate
for many of the objects.  One is a double match with 1H~1555-552, and
we remove the false match before presenting the table.  The objects
are listed in Table~\ref{HMXB_match}.  {The other matches are
  believed to be real, given that the magnitudes in the SPM catalog
  match well to the magnitudes in the HMXB catalog, and that the space
  density of such bright stars, and of X-ray sources is small.  For
  the two key objects of interest, we discuss the robustness of the
  matches to the SPM data.}

\begin{table}
\begin{tabular}{|l|r|r|r|r|r|r|r|r|r|r|r|r|}
\hline
  \multicolumn{1}{|c|}{Name} &
  \multicolumn{1}{c|}{RA} &
  \multicolumn{1}{c|}{DE} &
  \multicolumn{1}{c|}{$\mu_{RA}$} &
  \multicolumn{1}{c|}{$\mu_{DE}$} &
  \multicolumn{1}{c|}{$\epsilon_{RA}$} &
  \multicolumn{1}{c|}{$\epsilon_{DE}$} &
  \multicolumn{1}{c|}{SPMID} &
  \multicolumn{1}{c|}{P$_{orb}$} &
  \multicolumn{1}{c|}{SpType} &
  \multicolumn{1}{c|}{V} &
  \multicolumn{1}{c|}{$\ell$} &
  \multicolumn{1}{c|}{$b$} \\
\hline
  1WGA J0648.0-4419$^\dagger$ & 102.01958 & -44.31624 & -5.4 & 3.6 & 4.6 & 4.7 & 2800000062 & 1.55 & sdO6&8.3&254&-19\\
  3A 0726-260 & 112.22327 & -26.10800 & 0.2 & 3.0 & 1.9 & 1.7 & 5490005282 & 34.5 & O8-9Ve&11.9&240&-4\\
  1H 0739-529 & 116.84819 & -53.33218 & -7.2 & 38.4 & 2.2 & 2.3 & 1720000143 &  & B7 IV-Ve&8.6&266&-14\\
  4U 0900-40$^\dagger$ & 135.52858 & -40.55469 & -6.0 & 7.4 & 2.6 & 2.5 & 3450000153 & 8.96 & B0.5 Ib&6.9&263&+4\\
  GRO J1008-57 & 152.44569 & -58.29321 & -6.3 & 3.0 & 2.3 & 2.4 & 1350245141 & 135.0 & B0e&15.4&283&-2\\
  RX J1037.5-5647$^\dagger$ & 159.39714 & -56.79884 & -5.9 & 1.1 & 2.2 & 2.2 & 1780004465 &  & B0V-IIIe&11.5&285&+1\\
  1A 1118-615 & 170.23822 & -61.91672 & -4.1 & -5.5 & 2.4 & 2.7 & 1370072493 &  & O9.5Ve&12.2&292&-1\\
  4U 1119-603 & 170.31293 & -60.62380 & -8.7 & 9.2 & 2.7 & 3.0 & 1371719741 & 2.09 & O6.5 II-III&13.4&292&0\\
  IGR J11215-5952 & 170.44512 & -59.86333 & -5.0 & -2.7 & 3.9 & 3.9 & 1370003218 &  & B1Ia&10.0&292&+1\\
  IGR J11435-6109 & 176.04442 & -61.11709 & -10.1 & 7.3 & 2.2 & 2.3 & 1370120355 & 52.46 & B3e&13.4&295&+1\\
  2S 1145-619$^\dagger$ & 177.00010 & -62.20693 & -8.1 & 1.1 & 2.3 & 2.3 & 990000166 & 187.5 & B0.2IIIe&9.3&296&0\\
  1E 1145.1-6141 & 176.86901 & -61.95375 & -2.8 & 5.6 & 2.1 & 2.3 & 1371026060 & 14.4 & B2Iae&13.1&295&0\\
  4U 1223-624 & 186.65652 & -62.77034 & -1.6 & 1.6 & 0.8 & 0.9 & 1000002780 & 41.59 & B1-1.5 Ia&11.0&300&0\\
  1A 1246-588 & 192.40252 & -59.12163 & -7.8 & 8.5 & 2.9 & 3.1 & 1390161384 &  & 4U 1246-58&14.8&303&4\\
  1H 1249-637$^\dagger$ & 190.70944 & -63.05862 & -11.0 & -6.6 & 4.0 & 2.9 & 1010000094 &  & B0.5IIIe&5.3&302&0\\
  1H 1253-761$^\dagger$ & 189.81059 & -75.37058 & -23.8 & -11.2 & 4.5 & 4.9 & 400000126 &  & B7 Vne&6.5&302&-13\\
  1H 1255-567$^\dagger$ & 193.65370 & -57.16865 & -31.7 & -19.2 & 2.7 & 2.4 & 1390000175 &  & B5 Ve&5.1&303&+6\\
  4U 1258-61 & 195.32125 & -61.60186 & -0.7 & -1.5 & 1.4 & 1.6 & 1390063201 & 133.0 & B0.7Ve&14.3&304&+1\\
  4U 1538-52 & 235.59735 & -52.38599 & -6.5 & -6.3 & 1.6 & 1.8 & 1870167905 & 3.73 & B0 Iab&14.5&327&+2\\
  1H 1555-552 & 238.59090 & -55.32900 & -6.0 & 2.2 & 2.0 & 2.1 & 1870004373 &  & B2IIIn&8.8&327&-1\\
  IGR J16465-4507 & 251.64691 & -45.11794 & -0.8 & -4.5 & 3.8 & 4.1 & 3050085581 &  & B0.5I&14.8&340&0\\
  4U 1700-37$^\dagger$ & 255.98652 & -37.84420 & -0.4 & -2.8 & 2.5 & 2.7 & 3650000127 & 3.41 & O6.5Iaf&6.5&348&+2\\
  XTE J1739-302 & 264.79812 & -30.34372 & -5.6 & 16.1 & 5.8 & 5.7 & 5080066790 &  & O8Iab(f)&14.0&358&0\\
  RX J1744.7-2713 & 266.19070 & -27.22901 & 2.6 & 0.2 & 6.6 & 6.8 & 5080000512 &  & B0.5V-IIIe&8.4&1&1\\
  IGR J17544-2619 & 268.60533 & -26.33129 & -13.9 & 7.4 & 1.3 & 1.4 & 5801572773 &  & O9Ib&12.8&3&0\\
  SAX J1819.3-2525 & 274.83977 & -25.42719 & -0.7 & -6.8 & 6.6 & 6.5 & 5821604969 & 2.8 & B9III&14.4&7&-5\\
\hline
\end{tabular}
\caption{The proper motions of High Mass X-ray binaries in SPM 4.0.
  The columns are: (1) Name of the object, from the L07 catalog (2)
  right ascension, from SPM 4.0 (3) declination from SPM 4.0 (4)
  proper motion in right ascension, in milliarcseconds per year, from
  SPM 4.0 (5)proper motion in declination in milliarcseconds per year
  from SPM 4.0 (6) uncertainty on column 4 (7) uncertainty on column 5
  (8) SPM 4.0 source catalog number (9) orbital period in days from
  L07 (10) spectral type of the donor star from L07 (11) the $V$ band
  magnitude from the SPM catalog, (12) the Galactic longitude of the
  source rounded to the nearest degree and (13) the Galactic latitude
  of the source rounded to the nearest degree.  The Galactic
  coordinates are given to allow the reader an idea of the location
  within the Galaxy, and should not be used for matching with other
  catalogs.  Daggers indicate sources with more precise measurements
  from Hipparcos (Chevalier \& Ilovaisky 1998).  The source 1WGA
  J0648.0-4419 has a proper motion from Hipparcos, but is most likely
  an accreting white dwarf, rather than a high mass X-ray binary
  (e.g. Mereghetti et al. 2013).}
\label{HMXB_match}
\end{table}

We have cross-checked the proper motions for the brightest objects
against the Hipparcos catalog (Perryman et al. 1997; see also
Chevalier \& Ilovaisky 1998 who have already presented the Hipparcos
results for the high mass X-ray binaries).  Seven objects match within
0.5'' in the two proper motion catalogs, and for five of them, the
proper motions in both directions agree within 1$\sigma$, while one
(1H 1255-567) disagrees by 3.5 $\sigma$ in declination, and one
(4U~1700-37) disagrees by $2.8\sigma$ in declination.  These are two
extremely bright stars ($B$=4.99 and 6.74, respectively in the SPM
photometry), and saturation may have affected their positional
measurements.

The only high proper motion object which had not previously been
identified as such is IGR J17544-2619.  In the SPM data, it is seen
to have a proper motion of $-13.86\pm1.34$ masec/year in right
ascension, and $+7.31\pm1.41$ in declination.  The UCAC4 proper
motions are consistent with these values, but with larger errors:
$-17.1\pm5.5$ and $+10.7\pm1.9$ masec/yr in RA and Dec, respectively.
Finally we have inspected the Hipparcos versus SPM~4.0 data within
1 degree of this star, and have found a mean offset of $+2.5$
masec/year between the two fields in declination.  We take this to be
an estimate of the systematic errors of the SPM~4.0 data within this
crowded region near the Galactic Center, and take 2.5 mas/yr as an
uncertainty in both the RA and Dec directions -- it is likely that the
same effects of crowding led to the discrepancies in the proper
motions of the two sources discussed above for which we did not see
good agreement between SPM 4.0 and Hipparcos.

{We note that the probability that IGR J17544-2619 is a chance
  superposition is tiny.  The positions in Liu et al. (1997) and SPM
  match to within 0.05''.  Additionally, the source displays the X-ray
  source properties of the supergiant fast X-ray transients, and
  supergiant stars are both rare and bright.  The counterpart has also
  been verified to be a supergiant spectroscopically (Rahoui et
  al. 2008) and clearly shows radial velocity variations (Nikolaeva et
  al. 2013).}

\subsection{Low mass X-ray binaries}


We next consider the matches between low mass X-ray binaries from Liu
et al. (2006) and the SPM catalog.  A number of the globular cluster
X-ray sources match the SPM positions to within 2'', but these cannot
be considered reliable matches because of the high density of stars in
globular clusters.  Among the field stars, there are three matches
within 2'' of an X-ray source.  For the matches with separations
between 2'' and 5'', we check whether coordinate round-off may have
led to a larger separation in our matching process than the real
separation, but we find that these objects are all likely to be chance
superpositions.  As a result, we focus on the three matches that are
within 2''.

The three matching sources are GRO~J1655-40, GX 1+4 and 2A 1822-371.
The former two sources have proper motions consistent with zero in the
SPM data.  The proper motion of GRO~J1655-40 has been measured with
HST (Mirabel et al. 2002), and is statistically significantly
different from zero, but given the large measurement errors in the SPM
data for this object, the two results are consistent with one another.

The final object, 2A~1822-371, is of interest.  It shows a proper
motion in right ascension of $-12.11\pm2.18$ mas/year and in
declination of $8.55\pm2.14$ mas/year.  The data values are consistent
with the values of $-6\pm4$ and $14\pm4$, respectively from the UCAC4
catalog (Zacharias et al. 2012), but since the uncertainties are
smaller for the SPM catalog, we use the SPM values throughout the
paper.  No systematic offsets were found between the Hipparcos and SPM
measurements of nearby stars, and 2A~1822-371 itself is too faint
(roughly 15.2 in all filters from B through K) to have been detected
by Hipparcos.

\begin{figure}
\includegraphics[width=16 cm]{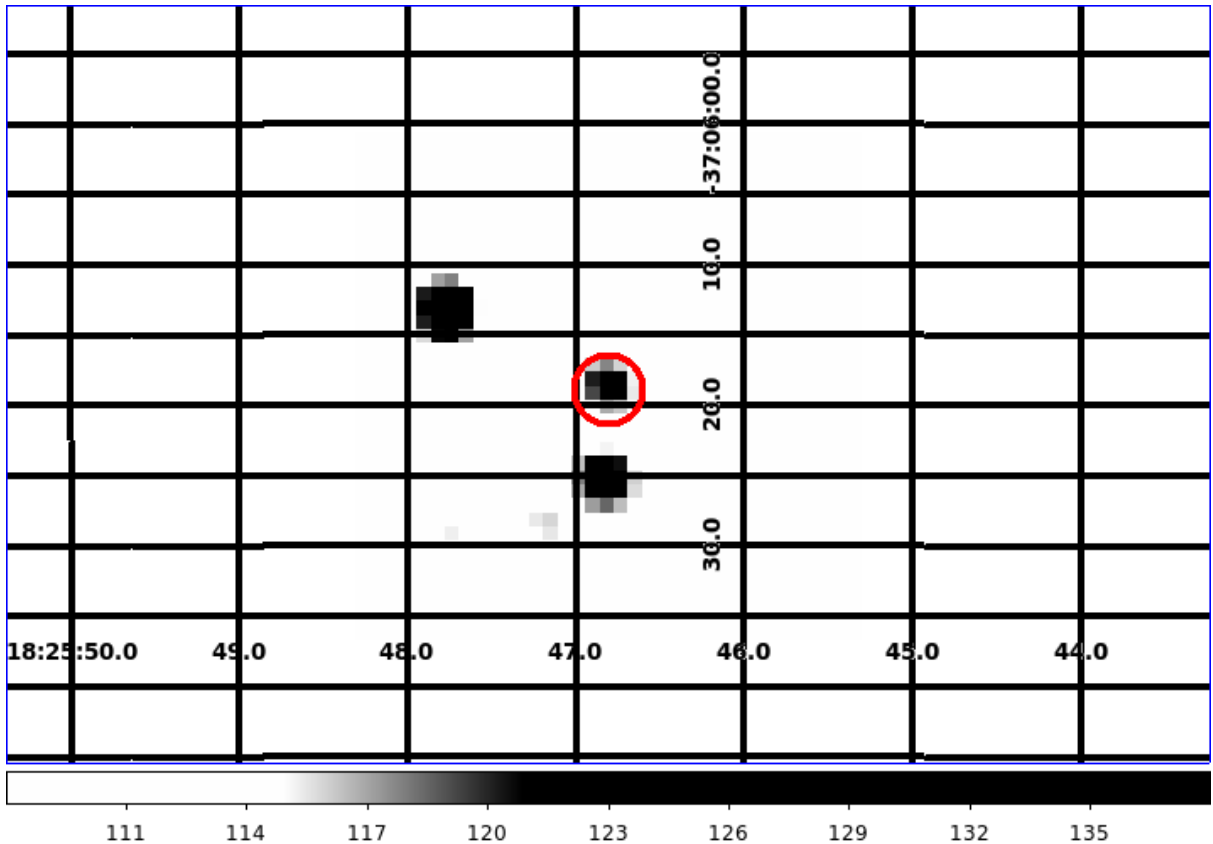}
\caption{The 2MASS image of 2A~1822-371.  The circle around the source
  is 2.5'' in radius, showing that a single star dominates the
  brightness within this region.  That star is clearly the counterpart
  of the X-ray source given that it shows the same periodicity in
  X-ray and optical data (Middleditch et al. 1981).}
\label{1822fig}
\end{figure}

The offset between the SPM and Liu et al (2006) positions for this
object is 0.5'', consistent with roundoff to the nearest 0.1'' in RA
and nearest 1.0'' in Declination.  The match is further understood to
be correct on the basis of the fact that the optical star and the
X-ray source show the same periodicity (the orbital period of the
binary).  An image of the field from 2MASS in $J$ band is shown in
figure \ref{1822fig}.\footnote{The 2MASS data offer the best digital
  optical/infrared data from Skyview.}

\section{Discussion: IGR J17544-2619}

The source IGR~J17544-2619 was first detected with INTEGRAL in a burst
of emission that lasted about two hours (Sunyaev et al. 2003), and is
a member of the class of objects called supergiant fast X-ray
transients (SFXTs).  The SFXTs are a class of objects discovered
mostly with INTEGRAL (Sguera et al. 2005) which show extremely bright
flares in the hard X-rays on timescales of minutes.  For an excellent
review on the topic, we refer the reader to Sidoli (2013), but we
briefly summarize some of the key properties of these systems.  About
half of the SFXTs show rapid periodicities consistent with the
rotation period of a neutron star (Sidoli 2013), and, by definition,
all members of the class have supergiant donor stars (e.g. Chaty
2010).  Many of the systems show modulation on the apparent orbital
periods of the systems (e.g. Sidoli et al. 2007).  In the Corbet
diagram of spin period plotted versus orbital period, the SFXTs share
paramter space with both the Be X-ray binaries and the traditional
persistent supergiant wind-fed systems (Sidoli 2013).  There remains
debate about whether the systems have stellar winds with strong
equatorial components (e.g. Sidoli et al. 2007), are clumpy (e.g. in
't zand 2005; Oskinova et al. 2012), or perhaps both.  Gated accretion
due to high magnetic fields and slow rotation periods is an additional
possibility (e.g. Grebenev \& Sunyaev 2007; Bozzo et al. 2008).

The full set of system parameters of IGR~J17544-2619 are well
described in Drave et al. (2012).  It has been shown to be associated
with the infrared counterpart 2MASS J17542527-2619526 (in 't Zand
2005).  Optical and infrared studies have further found that its donor
star is an O9Ib star, with a mass of 25-28 $M_\odot$ (Pellizza et
al. 2006) at a distance of about 3.6 kpc (Rahoui et al. 2008).  Its
orbital period has been identified by Clark et al. (2009) on the basis
of INTEGRAL timing to be 4.926$\pm$0.001 d, and its pulse period has
been found by Drave et al. (2012) to be 71.49$\pm$0.02 s.  Drave et
al. (2012) show that the system lies near the supergiant persistent
X-ray binaries in the Corbet diagram, and that the chief difference
relative to those systems is likely to be a higher eccentricity for
the binaries (see also in 't Zand 2005).

In light of the expectation that these systems might have large
eccentricities, measurements of their proper motions is especially
interesting, as the same processes that induce eccentricity in
previously circular binaries will tend to impart momentum to the
binary's center of mass.  These processes can include the loss of a
substantial amount of mass during the supernova explosion -- the
Blauuw kick -- and an asymmetric kick applied to the compact object
during or very shortly after the supernova explosion.

At least one runaway high mass X-ray binary is known, LS 5039 (Ribo et
al. 2002).  That system has a short orbital period, and a high
measured eccentricity, so it is quite similar in system parameters to
IGR~J17544-2619.  On the other hand, it is thought to have a more
rapidly rotating neutron star, such that the radio pulsar mechanism
works in that system, and its high energy emission is produced by a
shock between the stellar wind and the pulsar wind in the system
(e.g. Dubus 2006).

A useful starting point in understanding an X-ray binary system is to
determine whether its observed kinematical properties could be
explained only as the result of a Blauuw kick.  Nelemans et al. (1999)
work out the system velocity expected due to symmetric mass loss in a
supernovae explosion as:
\begin{equation}
v_{sys} = 213 \left(\frac{\Delta{M}}{M_\odot}\right) \left(\frac{m}{M_\odot}\right)\left(\frac{P_{re-circ}}{{\rm day}}\right)^{-\frac{1}{3}}\left(\frac{m+M_{rem}}{M_\odot}\right)^{-\frac{5}{3}} {\rm km/sec},
\label{Nelemans}
\end{equation}
in a form which is convenient, but appropriate only for low mass X-ray
binaries where there is sufficient time for the system to
re-circularize.  They had already derived that $P_{re-circ} =
P_{postSN}(1-e^2)^{3/2}$, which is more appropriate to the case of a
high mass X-ray binary.  We can then re-write equation \ref{Nelemans}
as:
\begin{equation}
v_{sys} = 213 \left(\frac{\Delta{M}}{M_\odot}\right) \left(\frac{m}{M_\odot}\right)\left(\frac{P_{postSN}}{{\rm day}}\right)^{-\frac{1}{3}}\left(1-e^2\right)^{-\frac{1}{2}}
\left(\frac{m+M_{rem}}{M_\odot}\right)^{-\frac{5}{3}} {\rm km/sec}
\end{equation}

Furthermore, $e = \frac{\Delta{M}}{M_{rem}+m}$ -- so it can be seen
immediately that for increasing $\Delta{M}$, both $e$ and $v_{sys}$
increase monotonically.  Furthermore, we can re-write the equation now
to remove $e$, getting:

\begin{equation}
v_{sys} = 213 \left(\frac{\Delta{M}}{M_\odot}\right) \left(\frac{m}{M_\odot}\right)\left(\frac{P_{postSN}}{{\rm day}}\right)^{-\frac{1}{3}}\left(1-\frac{\Delta{M}}{M_{0,t}}\right)^{-\frac{1}{2}}
\left(\frac{m+M_{rem}}{M_\odot}\right)^{-\frac{5}{3}} {\rm km/sec}
\label{symmetric}
\end{equation}
where $M_{0,t}$ is the initial total mass of the system.  In the case
of asymmetric kicks, the situation can be more complicated, since they
can counter-act the symmetric kicks (e.g. Kalogera 1998), but
generally speaking, large eccentricities and large space velocities
are expected to be seen in the same systems.

Starting from equation \ref{symmetric}, we can estimate the amount of
mass loss that must have taken place.  If we take the donor star mass
to be 25 $M_\odot$, and the accretor to be a 1.4$M_\odot$ neutron
star, we find that the systemic velocity will be about $13
\frac{\Delta{M}}{M_\odot}(1-e^2)^{-1/2}$ km/sec.  If the system is
assumed to be at a distance of 3.6 kpc, then its tangential velocity
relative to its Local Standard of Rest is about 275 km/sec out of the
Galactic Plane, {with a range of 160-320 km/sec for the rand of
possible distnaces from 2.1-4.2 kpc given in Rahoui et al. (2008).
The distance uncertainties dominate over the proper motion
uncertainties.}  Its mean systemic radial velocity is about 47 km/sec
(Nikolaeva et al. 2013).\footnote{Nikolaeva et al. (2013) also suggest
  that the orbital period may be about 12 days for this source, on the
  basis of 27 spectra.  This analysis gives a minimum mass for the
  compact object of 2.8 $M_\odot$, suggesting either that the neutron
  star is, by far, the heaviest neutron star measured to date, or,
  more likely, that the radial velocity curve has not yet been sampled
  well enough, given that fitting an eccentric orbit requires 6 free
  parameters.  We prefer the latter interpretation, largely on the
  basis of the very convincing X-ray modulation from the source found
  by Clark et al. (2009), and the finding that the masses of the
  neutron stars in high mass X-ray binaries tend to be well-clustered
  around 1.4 $M_\odot$ (Thorsett \& Chakrabarty 1999).  The mean
  radial velocity should be robust to uncertainties in the orbital
  solution.}

The system is thus just within the bounds possible for producing a
neutron star from a red supergiant without an asymmetric kick being
required (see e.g. Woosley et al. 2002).  The eccentricity, at least
before any tidal circularization processes have taken place, would be
about $13/39.4$, or about 1/3.  At the present time, no concrete
measurement exists for the eccentricity of this system.\footnote{If
  the Nikolaeva et al. (2013) results are taken at face value then the
  eccentricity is about 0.44.}  If the eccentricity is small, it may
be because the binary has become partly circularized.  Both in 't Zand
(2005) and Drave et al. (2012) have suggested that the major
difference between SFXTs and the more traditional persistent
supergiant X-ray binaries is that the SFXTs have higher orbital
eccentricities.  We thus do expect that the binary will not have been
completely circularized.

Some models of stellar evolution suggest that neutron stars can be
produced from progenitors with quite high initial masses that lose a
lot of mass in a Wolf-Rayet phase (Woosley et al. 2002).  If the
neutron star were produced from a massive star progenitor, which
underwent considerable mass loss in a Wolf-Rayet phase, then it is
likely that the progenitor was less than $14 M_\odot$ before the
supernova (Woosley et al. 2002), and that an asymmetric kick was
needed to give the system its observed proper motion.  Given the
considerable uncertainties, in mass loss rates from massive stars,
supernova explosions, and whether this system has mass transfer from
the progenitor star of the neutron star to the current donor star, it
is difficult to determine without detailed calculations that go beyond
the scope of this paper whether the system must have had a natal kick,
or whether all its properties can be explained by a simple impulsive
mass loss.

The essential point is that the large proper motion is in agreement
with the idea that the system at least formed with a large
eccentricity.  We can now look at circularization timescales of X-ray
binaries to make some estimate of for how long the eccentricity is
likely to have remained.  The theory of circularization of binaries is
generally tested by observations of open clusters (see e.g. Verbunt \&
Phinney 1995).  However, if, as suggested by Mathieu et al. (1992)
that binaries with such short orbital periods ordinarily circularize
on the pre-main sequence, then there is no empirical handle on how
quickly such systems should circularize.  Most theoretical work has
focused on the circularization of binaries with cooler stars
(e.g. Zahn 1977; Claret \& Cunha 1997; Goodman \& Dickson 1998), so
there is not currently even much theoretical work to use to test
whether significant circularization should take place in a mass
transferring binary with a supergiant donor star.

We can attempt to bound the system's age in a few ways.  First, we can
see if we can trace its motion back to any of the star forming regions
nearby in the catalog of Avedisova (2002).  The position matches up
well with that of region 3.270-0.102 but the distance for that star
forming region is given by Downes et al. (1980) to be 18 kpc, so the
association is unlikely.  In the absence of a birth site, we can make
a crude estimate of the age, on the basis that the system is moving
away from the Galactic Plane at about 10 msec/year, and is unlikely to
have formed more than about 50 pc from the Galactic Plane.  This
yields an age of a few hundred thousand years, so if substantial
circularization has taken place, then we immediately learn something
interesting about circularization of supergiant binaries.  On the
other hand, if we take its high eccentricity as a given based on its
SFXT nature, we can use it as a probe of circularization in binaries.
Then we can see that it is moving toward the nearest supernova
remnant, and is located several degrees from the next nearest
supernova remnant (Green 2009), making an association with a supernova
remnant unlikely and thus making the system age likely to be larger
than the typical lifetimes of supernova remnants.  It is thus unlikely
to be much younger than about $10^4$ years.  Further searches for
associations (e.g. from the Vista Variables in the Via Lactea project
-- Minniti et al. 2010) with young star associations may potential
help locate the birth site of this X-ray binary.

\subsection{Runaway high mass X-ray binaries and the short GRB problem}

The short gamma-ray bursts are thought to be produced by mergers of
compact objects, either between two neutron stars or between a neutron
star and a black hole (e.g. Paczynski 1986).  As such, high mass X-ray
binaries represent excellent candidate progenitors for these events.
Furthermore, many of these short gamma-ray bursts have been located
not in the centers of galaxies, near the location of maximum star
formation, but rather in the outskirts of galaxies (e.g. Gehrels et
al. 2005).  The discovery of at least one such object with a peculiar
velocity of the same order as the velocity of the Galactic rotation
curve helps point the way toward finding the actual progenitors of the
subclass of short GRBs taking place outside of the main bodies of
galaxies.  In particular, the second velocity kick that will be
applied at the time of the second supernova explosion could easily be
large enough to unbind the system from the Galaxy without unbinding
the binary itself.

The cleanest sample of high mass X-ray binaries whose kinematic
properties are well understood is the Hipparcos sample (Chevalier \&
Ilovaisky 1999 -- CI99).  Hipparcos made measurements of 17 HMXBs,
including 4 with supergiant donor stars and 13 with Be donor stars (CI
99).  The four supergiant systems are all found to have peculiar
tangential velocities of less than 100 km/sec.  The largest quoted
tangential velocity is that of Cygnus X-1; the proper motion of Cyg
X-1, is, in fact, very similar to that of the Cygnus OB3 association,
so that it is unlikely to have formed with a large velocity kick
(Mirabel \& Rodrigues 2003).  The other high mass X-ray binaries
studied with Hipparcos have tangential velocities that are larger than
the typical velocity dispersion of massive stars, but considerably
smaller than the Galactic escape velocity, or even the Galactic
rotational velocity.  The Be X-ray binaries measured by Hipparcos have
even lower system tangential velocities -- typically about 10 km/sec.
That the Be X-ray binaries were formed with relatively small velocity
kicks is not surprising given that very large kicks would have unbound
most of them.

Developing an understanding of these objects is important, as well,
for understanding where gravitational wave sources are likely to be
located and determining strategies for conducting electromagnetic
follow-up of them.  At the present time, it is not clear what will be
the eventual fate of IGR J17544-2619 -- but with continued improvement
in understanding its orbital parameters it should be possible to
conduct binary evolution simulations (e.g. Belczynski et
al. 2006;2008) of it to estimate the probability it will eventually
turn into a double neutron star which merges in less than a Hubble
time.

\subsection{Linear momentum}
{
The linear momentum of the system has a large uncertainty due to the
distance uncertainty.  Taking the best estimate distance and the best
estimate for the systemic radial velocity, the space velocity of the
system is about 280 km/sec.  The donor star mass has been estimated to
be 25 $M_\odot$ (Rahoui et al. 2008), so the systemic mass should be
about 26-27 $M_\odot$.  The total momentum is then equivalent to that
of a pulsar with a 5300 km/sec space velocity.  This value is far in
excess of the largest space velocities seen from isolated pulsars
(Hobbs et al. 2005).  It thus requires that either there is a
population of pulsars which has not been detected because pulsars born
with such a high velocity quickly escape the Galaxy, or, more likely,
that the kick comes primarily from the Blaauw mechanism.  If the kick
comes primarily from the Blauuw mechanism, then we can establish that
there must have been dramatic mass loss during the supernova that
produced the neutron star.  Given that the system must also be quite
young, an interesting implication is that the donor star may still be
polluted with supernova ejecta.  Abundance studies of the donor stars
in SFXTs may thus be an interesting topic for future research.}

\section{Discussion: 2A 1822-371}

The source 2A~1822-371 differs from the bulk of low mass X-ray
binaries in several ways.  First, it is a slow pulsar, with a pulse
period of 0.59 seconds (Jonker et al. 2001).  Nearly all the low mass
X-ray binaries in the Galaxy with known spin periods have spin periods
less than 10 milliseconds (see Liu et al. 2007).  The ones which have
longer measured spin periods are the ultracompact X-ray binary
4U~1626-67 (Middleditch et al. 1981), the symbiotic X-ray binary GX
1+4, the intermediate mass X-ray binary Her X-1 (Tananbaum et
al. 1972; Crampton 1974) and the ``Bursting Pulsar'' (Finger et
al. 1996).  None of these systems is a traditional low mass X-ray
binary with a main sequence (or mildly evolved) donor star of less
than 1 $M_\odot$ except 2A~1822-371.

That so few X-ray binaries are seen with slow pulsations, main
sequence donors and orbital periods of 3-10 hours is not surprising.
Standard theories of the evolution of neutron stars' spin periods and
magnetic fields predict that only a small fraction of the donor star's
mass much be accreted in order to spin up the neutron star into a
millisecond pulsar (e.g. Alpar \& Shaham 1985).  The slow pulsar phase
of an X-ray binary's lifetime is therefore expected to be a short one.

\subsection{Position in the Galaxy}

A variety of approaches have been used to estimate the distance to
2A~1822-371.  Because the system is persistent, one cannot use the
standard method of estimating the flux and temperature of the donor
star in quiescence.  Because the system is a pulsar and does not show
Type I bursts, the bursts cannot be used as standard candles.
Instead, we rely on the work of Mason \& Cordova (1982), who modelled
the disk rim emission from this nearly edge-on system, and found that
the system's distance is in the 1-5 kpc range, depending on the exact
assumptions used.

The position of 2A~1822-371 is 18:25:46.81 -37:06:18.6 (Cutri et
al. 2003) -- in Galactic coordinates, this corresponds to 356.8502,
-11.2908.  If we trace the proper motion backwards from the present
time, it intersects the Galactic Plane in about 3.2 Myrs, at almost
exactly the projected location of the Galactic Center.  As a result,
given that the Galaxy has a nearly constant rotational velocity as a
function of Galactocentric radius outside the very inner Bulge, one
can take its proper motion to be the same as its initial velocity
kick.\footnote{Given the very large uncertainties on the system's
  distance and the susbtantial uncertainties on the proper motions, we
  simplify the calculations to ignore its acceleration in the Galactic
  potential, which will, in any even have a small effect in a few
  Myrs.}  We note that the object could not have formed {\it in} the
Galactic Center region 3.2 Myr ago unless its distance has been badly
mis-estimated -- Jonker et al. (2003) find a mean radial velocity of
only about 54$\pm$24 km/sec, so that in 3.2 Myrs, the distance
travelled, ignoring acceleration, would be only a few hundred pc, not
enough to bring a source from the Galactic Center to the upper bound
distance of 5 kpc.

{Finally, we estimate the space velocities for the source, in
  Galactic coordinates, where $U$ is the velocity in the direction of
  the Galactic anti-center, $V$ is the velocity along the direction of
  the Galactic rotation curve, and $W$ is the velocity in the
  direction toward the North Galactic Pole.  These total velocity is
  about 100 km/sec if the source is at a distance of 1 kpc, and about
  365 km/sec if the source is at a distance of 5 kpc; these distances
  represent the range of possible distances given in Mason \& Cordova
  (1982).  In either case, the tangential motion is directly away from
  the Galactic Center.
}

\subsection{Age and eccentricity connection?}

A coherent picture can be developed based on the idea that 2A~1822-371
is only about 3.2 Myrs old.  In such a case, the system may still be a
mildly eccentric binary in Roche lobe contact.  Hut \& Paczynski
(1984) showed that even very small eccentricities of Roche lobe
overflowing systems could lead to large changes in the mass accretion
rate -- essentially, because the scale height of a stellar atmosphere
for a typical main sequence star is about $10^{-4}$ times its radius,
a change of a factor of a few in mass transfer rate should result from
an eccentricity of order $10^{-4}$.  Such changes would be nearly
impossible to measure from X-ray binaries except possibly through
careful pulse timing.  For 2A~1822-371, the best observational
constraint on the eccentricity requires only that it be less than 0.03
(Jonker \& van der Klis 2001).

It is generally expected that X-ray binaries (and cataclysmic
variables) with orbital periods less than about 4 hours will have
their orbital evolution, and hence their mass transfer rate,
determined mostly by the effects of gravitational radiation (Kraft et
al. 1962).  Interestingly, 2A~1822-371 shows two separate lines of
evidence that its mass transfer rate is much higher than would be
expected from gravitational radiation.  It is luminous in the X-rays,
despite being an accretion disk corona source, in which one expects
only about 1\% of the emission to be scattered into the observer's
line of sight; and it shows a large, positive orbital period
derivative from eclipse timing, suggesting super-Eddington,
non-conservative mass transfer, where a mass transfer rate about
$10^3$ times lower would be expected for a circular orbit at that
period (Burderi et al. 2010).  Invoking an eccentricity of order
$10^{-3}$ would allow for the high mass transfer rate, while also not
producing any directly detectable signature of the eccentricity.  An
eccentricity that large should be produced in a low mass X-ray binary
at birth in all cases -- even in the extreme case of an accretion
induced collapse of a white dwarf would result in a large enough loss
of mass-energy as the difference in binding energy of the compact
object.  This would not give the $\sim100-500$ km/sec space velocity we
observe for 2A~1822-371, but the mass loss from a pre-SN core of about
$3M_\odot$ would.

Maccarone (2005) summarized a range of literature on the
circularization of close binaries.  Because short period binaries in
open clusters appear to be circularized on the pre-main sequence
(Mathieu et al. 1992), it is difficult to approach empirically the
problem of the circularization of binaries which reach short periods
as the result of common envelope evolution and then become eccentric
as the result of supernovae.  Goodman \& Dickson (1998) estimate the
circularization timescale for solar-like binaries to be about 150 Myr
$P_{orb}^{3}$, with $P_{orb}$ being the orbital period in days.  In
such a case, the circularization timescale would be about 1 Myr for
2A~1822-371.  Since the inferred eccentricity at formation would be
$\sim0.3$ if $\sim1/3$ of the mass of the progenitor binary were lost,
the system could be $\sim10$ circularization timescales old, and still
have $10^{-3}$ as its current eccentricity.  If resonant effects
(Witte \& Savnoije 2002) and mass transfer (e.g. Claret \& Cunha 1997)
speed up the circularization process dramatically, even the $10^{-3}$
eccentricty we have invoked may be too large, but at the present time,
the scenario seems reasonable, albeit not proved beyond doubt by the
present data.  Better timing of the pulsar, as may be possible with
LOFT (Feroci et al. 2012) could provide better constraints on the
eccentricity of the orbit.

\subsection{An asymmetric kick versus a Blauuw kick?}

In some cases, the combination of the three-dimensional velocity of a system
and its inclination angle can give information about the relative strengths of
the Blaauw kick versus an asymmetric impulse.  This is most clear in the case
of a face-on binary, for which any large radial velocity offset from the local
standard of rest must be due to an asymmetric kick.  In the case of
2A~1822-371, the system is nearly edge on, and the velocity difference from
the local standard of rest is mostly in the tangential direction (Jonker et
al. 2003).  In this case, the possibility that the kick is dominated by the
Blaauw mechanism cannot be discarded out of hand, but does require a fine
tuning of the orbital phase at the time of the supernova explosion.  With a
single system, no definitive statements can be made; however, if Gaia measures
large proper motions relative to the local standard of rest for a large sample
of nearly edge-on X-ray binaries, it would be possible to make a statistical
statement from such a result.

\subsection{Linear momentum}
{
The linear momentum of the system has a large uncertainty due to the
distance uncertainty.  Taking the best estimate distance and the best
estimate for the systemic radial velocity, the space velocity of the
system is about 100 km/sec.  The donor star mass has been estimated to
be 0.4 $M_\odot$ (Cowley et al. 2003), so the systemic mass should be
about 1.8 $M_\odot$.  The total momentum is then equivalent to that of
a pulsar with a 150 km/sec space velocity.  This value is well within
the range of values for isolated radio pulsars (see e.g. Hobbs et
al. 2005, who find that a 265 km/sec Maxwellian is a good description
of the space velocity idstribution of pulsars).
}

\begin{table}
\begin{tabular}{llll}
\hline
Source & U & V & W\\
\hline
2A~1822-371 (1 kpc) &-76 &21 &64 \\
2A~1822-371 (5 kpc) &-132 &65 &335 \\
IGR J17544-2619 (3.6 kpc)&-57 &4 & 273 \\
IGR J17544-2619 (2.1 kpc)&-57 &9 & 162 \\
IGR J17544-2619 (4.2 kpc)&-58 &2 & 317 \\
\hline

\end{tabular}
\caption{The space velocities, in km/sec, in Galactic coordinate components for the two sources with newly identified large proper motions.  The values for 2A~1822-371 are given for distances of 1 and 5 kpc, and the values for IGR~J17544-2619 are given for distances of 3.6, 2.1 and 4.2 kpc.}
\label{vel_table}
\end{table}

\section{Summary}

We have discussed the matches between the catalogs of bright X-ray
binaries and the Yale Southern Proper Motion Survey 4.0 catalog.  We
have found two interesting binaries which are both unusual in their
X-ray properties, and show large proper motions away from the Galactic
Plane, and we have tabulated their three-space velocities in Table
\ref{vel_table}.  In both cases, the X-ray data are suggestive of
eccentricities created at the same time the systems received their
large peculiar velocities.

\section{Acknowledgments}
TJM thanks the Astrophysics Institute of the Canary Islands for
hospitality while a portion of this work was done.  He also thanks
Dave Russell, Tariq Shahbaz and Jorge Casares for useful discussions
about the system parameters of X-ray binaries, and Tony Bird and Seb
Drave for useful discussions about supergiant fast X-ray transients.
We also thank Paul Sell for a critical reading of the manuscript which
has improved its clarity and rigor substantially.

\label{lastpage}

\end{document}